\documentclass[PAPER,  cernpreprint,  atlasdraft=false,  coverpage=false,  texlive=2023, UKenglish, texmf, orcidlogo]{atlasdoc}
 
\PreprintIdNumber{CERN-EP-2022-052}

\usepackage{atlasphysics}
\usepackage{atlasheavyion}
\usepackage{newtxmath}
\usepackage{atlaspackage}
\usepackage{atlasbiblatex}

\usepackage{atlasphysics}

\usepackage[backend=biber]{biblatex}
\graphicspath{{logos/}{figures/}}

\usepackage{amsmath}
\usepackage{bm}
 
\newcommand{\etfcal}{\Sigma E_{\mathrm{T}}}
\newcommand{\nchrec}{N_\mathrm{ch}^\mathrm{rec}}
\newcommand{\Deta}{\mbox{$\Delta \eta$}}
\newcommand*{\HIJING}{\textsc{Hijing}\xspace}
\newcommand{\ptk}[1] {p_{\mathrm{T,#1}}}
\newcommand{\rhon}[1] {\rho_{#1}}
\newcommand{\cov}[1] {\mathrm{cov}_{#1}}
\newcommand{\var}[1] {\mathrm{var}(v_{ #1 }^2)}
\newcommand{\bq}{\textbf{q}}

\newcommand{\lr}[1]{\left\langle #1\right\rangle}
\newcommand{\llrr}[1]{\left\langle\left\langle#1\right\rangle\right\rangle}

\AtlasTitle{Correlations between flow and transverse momentum in Xe+Xe and Pb+Pb collisions at the LHC with the ATLAS detector: a probe of the heavy-ion initial state and nuclear deformation}

\AtlasAbstract{
The correlations between flow harmonics $v_n$ for $n=2$, 3 and 4 and mean transverse momentum~$[p_\mathrm{T}]$ in $^{129}$Xe+$^{129}$Xe and $^{208}$Pb+$^{208}$Pb collisions at $\sqrt{s_{_\mathrm{NN}}}=5.44$~TeV and 5.02~TeV, respectively, are measured using charged particles with the ATLAS detector. The correlations are potentially sensitive to the shape and size of the initial geometry, nuclear deformation, and initial momentum anisotropy. The effects from nonflow and centrality fluctuations are minimized, respectively, via a subevent cumulant method and an event-activity selection based on particle production at very forward rapidity. The $v_n$--$[p_\mathrm{T}]$ correlations show strong dependencies on centrality, harmonic number $n$, $p_{\mathrm{T}}$ and pseudorapidity range. Current models qualitatively describe the overall centrality- and system-dependent trends but fail to quantitatively reproduce all features of the data. In central collisions, where models generally show good agreement, the $v_2$--$[p_\mathrm{T}]$ correlations are sensitive to the triaxiality of the quadruple deformation. Comparison of the model with the Pb+Pb and Xe+Xe data confirms that the $^{129}$Xe nucleus is a highly deformed triaxial ellipsoid that has neither a prolate nor oblate shape. This provides strong evidence for a triaxial deformation of the $^{129}$Xe nucleus from high-energy heavy-ion collisions.}
 
\author{The ATLAS Collaboration}
 
\AtlasRefCode{HION-2021-01}
 
\PreprintIdNumber{CERN-EP-2022-052}

\AtlasHeading{ATLAS Paper}

\AtlasJournalRef{Phys. Rev. C 107 (2023) 054910}
\AtlasDOI{10.1103/PhysRevC.107.054910}

\hypersetup{pdftitle={ATLAS document},pdfauthor={The ATLAS Collaboration}}
\begin{document}
\maketitle
\section{Introduction}
\label{sec:1}
Heavy-ion collisions at the Relativistic Heavy Ion Collider (RHIC) and the Large Hadron Collider (LHC) produce quark--gluon plasma (QGP) whose space-time evolution is well described by relativistic viscous hydrodynamics~\cite{Gale:2013da,Heinz:2013th,Romatschke:2017ejr}. Driven by the large pressure gradients, the QGP expands rapidly in the transverse plane, and converts the spatial anisotropy in the initial state into momentum anisotropy in the final state. The collective expansion in each event is quantified by Fourier expansions of particle distributions in azimuth given by $dN/d\phi = (N/2\pi) (1+2\sum_{n=1}^\infty v_n \cos\,n(\phi-\Psi_n))$, where $v_n$ and $\Psi_n$ represent the amplitude and phase of the $n^{\mathrm{th}}$-order azimuthal flow vector, ${\bm V}_n=v_n{\textrm e}^{{\textrm i}n\Psi_n}$. The ${\bm V}_n$ are determined by the hydrodynamic response to the initial spatial anisotropy, characterized by eccentricity vectors ${\mathcal{E}}_n=\varepsilon_n{\textrm e}^{{\textrm i}n\Phi_n}$~\cite{Gardim:2011xv,Gale:2012rq}. Model calculations show that the ${\bm V}_n$ values are approximately proportional to $\mathcal{E}_n$ for $n=2$ and 3, as well as for $n=4$ in central collisions~\cite{Teaney:2010vd,Gardim:2011xv,Niemi:2012aj}. The measurements of $v_n$ and $\Psi_n$~\cite{Adare:2011tg,ALICE:2011ab,Aad:2012bu,Chatrchyan:2013kba,Aad:2013xma,Aad:2014fla,Aad:2015lwa} have placed important constraints on the properties of the medium and on the initial-state density fluctuations~\cite{Luzum:2012wu,Teaney:2010vd,Gale:2012rq,Niemi:2012aj,Qiu:2012uy,Teaney:2013dta} in high-energy nuclear collisions.
 
In addition to generating anisotropic flow, the hydrodynamic response to the fluctuations in the overall size of the overlap region also leads to fluctuations in the ``radial flow'', reflected by the average transverse momentum of particles in each event, $[\pT]$.\footnote{ATLAS uses a right-handed coordinate system with its origin at the nominal interaction point (IP) in the center of the detector and the $z$-axis along the beam pipe. The $x$-axis points from the IP to the center of the LHC ring, and the $y$-axis points upward. Cylindrical coordinates $(r,\phi)$ are used in the transverse plane, $\phi$ being the azimuthal angle around the beam pipe. The pseudorapidity is defined in terms of the polar angle $\theta$ as $\eta=-\ln\tan(\theta/2)$.} In particular, events with similar total energy but smaller transverse size in the initial state are expected to have a stronger radial expansion and therefore a larger $[\pT]$~\cite{Bozek:2012fw,Bozek:2017jog}. Furthermore, correlations between the ${\mathcal{E}}_n$ and the size in the initial state are expected to generate dynamical correlations between $v_n$ and $[\pT]$ in the final state. A Pearson coefficient has been proposed to study these correlations~\cite{Bozek:2016yoj},
\begin{align} \label{eq:rho}
\rhon{n}  =\frac{\llrr{v_n^2\delta\pT}}{\sqrt{\left(\lr{v_n^4}-\lr{v_n^2}^2\right)}\sqrt{\llrr{\delta \pT\delta \pT}}}\;,
\end{align}
where $\delta \pT = \pT-[\pT]$,  the ``$\llrr{}$'' denotes averaging over all particle pairs or triplets for events with comparable particle multiplicity, and the ``$\left\langle\right\rangle$'' denotes an averaging over events.   dependence of collision geometry and nucleon fluctuations,  the $\rhon{n}$ is expected to be positive in central and mid-ceDue to centralityntral collisions and negative in the peripheral collisions~\cite{Schenke:2020uqq}.  In the presence of initial momentum anisotropy,  the $\rhon{n}$ could turn into positive in the most peripheral region~\cite{Giacalone:2020byk}.  The  root-mean-square size of the nucleon also influences the behavior of $\rhon{n}$ in peripheral  collisions,  e.g.  a larger nucleon size would decrease the value of $\rhon{n}$~\cite{Giacalone:2021clp}.  ATLAS published a measurement of $\rhon{n}$ for $n=2$, 3 and 4 in Pb+Pb collisions at $\sqn=5.02$~\TeV~\cite{Aad:2019fgl},  which was followed by a similar measurement from ALICE~\cite{ALICE:2021gxt} in Pb+Pb and Xe+Xe collisions.  The results show positive correlations for all harmonics, except in the peripheral region, where $\rhon{2}$ is negative.  These behaviors have been qualitatively reproduced by recent initial-state model and hydrodynamic model calculations~\cite{Schenke:2020uqq,Giacalone:2020dln}.
 
Recent studies show that the $v_n$, $[\pT]$ and $v_n$--$[\pT]$ correlations in central collisions are also sensitive to the shape of atomic nuclei~\cite{Giacalone:2019pca,Giacalone:2020awm,Jia:2021tzt,Jia:2021wbq,Bally:2021qys,Jia:2021qyu}.  Most nuclei are more or less deformed into an ellipsoidal shape, for which the nuclear surface of the nucleon distribution can be described by~\cite{bohr},
\begin{align}\label{eq:R}
R(\theta,\phi) = R_0\left(1+\beta[\cos \gamma Y_{2,0}+ \sin\gamma Y_{2,2}]\right),
\end{align}
where $R_0$ is the nuclear radius, $Y_{l,m}$ are spherical harmonics, $\beta$, and $\gamma$ are quadrupole deformation parameters. The parameter $\beta$ is the magnitude of the deformation, with typical values of 0.1--0.4~\cite{Moller:2015fba}, while the angle $\gamma$, in the range $0\leq\gamma\leq 60^{\circ}$, describes the length imbalance of the three semi-axes $r_1,r_2,r_3$ of the ellipsoid, also known as triaxiality. The values $\gamma=0^{\circ}$, $\gamma=60^{\circ}$, and $\gamma=30^{\circ}$ correspond to the prolate ($r_1=r_2<r_3)$, oblate ($r_1<r_2=r_3$) and maximum triaxiality ($2r_2=r_1+r_3$) cases. Traditionally, the shapes of nuclei are inferred from low-energy spectroscopic measurements, which determine the shape parameters $(\beta,\gamma)$ for even--even nuclei such as $^{208}$Pb~\cite{Raman:1201zz}. The shape of odd-mass nuclei such as $^{129}$Xe can only be calculated using nuclear structure models that have been tuned to describe the even--even nuclei data. In this sense, flow measurements in high-energy heavy-ion collisions serve as a new tool to probe the nuclear shape, in particular for odd-mass nuclei. Recent model studies show that the $v_2$ and $\rhon{2}$ follow a simple parametric form~\cite{Giacalone:2018apa,Jia:2021tzt,Jia:2021qyu},
\begin{align}\label{eq:form}
v^2_2 \approx a+b\beta^2\;,\;\rhon{2}\approx a'+b' \cos(3\gamma)\beta^3\;.
\end{align}
The parameters $a$ and $a'$ represent values for collisions of spherical nuclei. They are the smallest in central collisions, whereas the parameters $b$ and $b'$ are nearly independent of centrality. Therefore, the impact of nuclear deformation is expected to be largest in central collisions. A large quadruple deformation for $^{129}$Xe of $\beta_{\mathrm{Xe}}\approx$~0.16--0.2 was extracted from the enhanced ratio $v_{2,\mathrm{Xe}}/v_{2,\mathrm{Pb}}$ in central collisions~\cite{Acharya:2018ihu,Sirunyan:2019wqp,Aad:2019xmh}. The measurement of $\rhon{2}$ here can then be used to further constrain the triaxiality of $^{129}$Xe.
 
This paper studies the centrality and system-size dependences of $\rhon{n}$ in $^{129}$Xe+$^{129}$Xe and $^{208}$Pb+$^{208}$Pb collisions to shed light on the effects of initial-state geometry  and nuclear deformation. The measurements are performed in several ranges of $\pT$ and $\eta$ to quantify the influence of final-state effects~\cite{Schenke:2020uqq}. The $\rhon{n}$ values are also influenced by nonflow effects from resonance decays and jets, which can be suppressed using the ``subevent method''~\cite{Jia:2017hbm,Huo:2017nms}, where the correlations are constructed by using particles from different subevents separated in $\eta$. The previous ALICE measurement of $\rhon{2}$ in Xe+Xe collisions~\cite{ALICE:2021gxt} was performed in wide centrality ranges with limited statistical precision. The larger acceptance of the ATLAS detector and a factor of ten more Xe+Xe events enable more precise measurements of $\rhon{2}$ in finer centrality ranges. A comparison of results in Xe+Xe and Pb+Pb collisions and model predictions then provides insight into the nuclear deformation and the nature of the initial sources responsible for harmonic flow and radial flow.
 
This paper also explores the issue of ``centrality fluctuations'', which refers to the fact that an experimental centrality definition based on the final-state particle multiplicity in an $\eta$ range is subject to smearing due to fluctuations in the particle production process. Such centrality fluctuations, also known as volume fluctuations~\cite{Skokov:2012ds,Luo:2013bmi}, have been shown to affect flow fluctuations~\cite{Zhou:2018fxx,Aaboud:2019sma} and $\rho_n$ values~\cite{Bozek:2020drh,Schenke:2020uqq,Jia:2021wbq}. This analysis explores the influence of centrality fluctuations on $\rhon{n}$ using two reference event-activity estimators: the total transverse energy $\Sigma \eT$ in the forward pseudorapidity range $3.2<|\eta|<4.9$ and the number of reconstructed charged particles $\nchrec$ in the mid-rapidity range $|\eta|<2.5$. Previous measurements of flow fluctuations show that $\Sigma \eT$ has better centrality resolution than $\nchrec$~\cite{Aaboud:2019sma}. This conclusion was also reached by model investigations of the forward--backward multiplicity correlation in Pb+Pb collisions~\cite{Jia:2020tvb,Yousefnia:2021cup}. Therefore, the default results are obtained using $\Sigma \eT$, while those based on $\nchrec$ give a sense of the extent of centrality fluctuations.
 
The paper is organized as follows. Sections~\ref{sec:2} and \ref{sec:3} describe details of the detector, event, and track selections. Section~\ref{sec:4} introduces the observables and subevent methods used in this analysis. The correlation analysis and systematic uncertainties are described in Sections~\ref{sec:5} and~\ref{sec:6}, respectively. Section~\ref{sec:7} presents the results of $\rhon{n}$ in the two collision systems, and discusses the role of nonflow and centrality fluctuations. Section~\ref{sec:8} compares the results with model predictions. A summary is given in Section~\ref{sec:9}.
\section{ATLAS detector and trigger}
\label{sec:2}
The ATLAS detector~\cite{Aad:2008zzm} provides nearly full solid-angle coverage with tracking detectors, calorimeters, and muon chambers, and is well suited for measurements of multiparticle correlations over a large pseudorapidity range. The measurements are performed using the trigger system, the inner detector~(ID), the forward calorimeters~(FCal), and the zero-degree calorimeters (ZDC). The ID detects charged particles within $|\eta| < 2.5$ using a combination of silicon pixel detectors, silicon microstrip detectors~(SCT), and a straw-tube transition-radiation tracker, all immersed in a 2~T axial magnetic field. The FCal consists of three sampling layers, longitudinal in shower depth, and covers $3.2<|\eta|< 4.9$. The ZDC are positioned at $\pm$140~m from the IP, detecting neutrons and photons with $|\eta|>8.3$. An extensive software suite~\cite{ATL-SOFT-PUB-2021-001} is used in the reconstruction and analysis of real and simulated data, in detector operations, and in the trigger and data acquisition systems of the experiment.

The ATLAS trigger system~\cite{Aaboud:2016leb} consists of a level-1 (L1) trigger implemented in dedicated electronics and programmable logic, and a high-level trigger (HLT) which uses software algorithms similar to those applied in the offline event reconstruction. During Xe+Xe data-taking, the minimum-bias trigger selected events with either more than 4~\GeV\ of transverse energy recorded in the whole calorimeter system at L1 ($\eT^{\mathrm{L1}}$) or a reconstructed track with $\pT>0.2$~\GeV\ at the HLT. In the Pb+Pb data-taking, the minimum-bias trigger required either $\eT^{\mathrm{L1}}>50$~\GeV\ or the presence of at least one neutron on both sides of the ZDC and a track identified by the HLT. To enhance the number of recorded events for ultracentral  Pb+Pb collisions, a dedicated trigger selected on the $\eT^{\mathrm{L1}}$ and the total transverse energy in the FCal, $\Sigma \eT$, at the HLT. The combined trigger selects events with $\Sigma \eT$ larger than one of the three threshold values: 4.21~\TeV, 4.37~\TeV, and 4.54~\TeV. The ultracentral trigger has a sharp turn-on as a function of $\etfcal$, and for these thresholds the trigger is fully efficient for the 1.3\%, 0.5\%, and 0.1\% of events in centrality percentile to be defined below, respectively. The fraction of events containing more than one inelastic interaction (pileup) is around $0.003$ in Pb+Pb data and around $0.0002$ in Xe+Xe data.
\section{Event and track selection}
\label{sec:3}
The analysis is based on ATLAS datasets corresponding to integrated luminosities of 3~$\mu\textrm{b}^{-1}$ of minimum-bias Xe+Xe data recorded at $\sqn = 5.44$~\TeV\ in 2017 and 22 $\mu\mathrm{b}^{-1}$ of minimum-bias and 470~$\mu\mathrm{b}^{-1}$ of ultracentral Pb+Pb data recorded at $\sqn = 5.02$~\TeV\ in 2015. The offline event selection requires a reconstructed primary vertex with its $z$ position satisfying $|\zvtx|< 100$~mm. For the Pb+Pb dataset, pileup events are suppressed by exploiting the expected correlation between the estimated number of neutrons in the ZDC and $\nchrec$. For both the Pb+Pb and Xe+Xe datasets, a requirement is also imposed on the correlation between $\etfcal$ and $\nchrec$ to further reduce the number of background events. The events are classified into centrality intervals based on the $\etfcal$ in the FCal. A Glauber model~\cite{Miller:2007ri,Aad:2011yr} is used to parameterize the $\etfcal$ distribution and provide a correspondence between the $\etfcal$ distribution and the sampling fraction of the total inelastic Pb+Pb or Xe+Xe cross-section, allowing centrality percentiles to be set. This analysis is restricted to the 0--84\% most central collisions, where the triggers are fully efficient and the contamination from photonuclear processes is small~\cite{Aad:2019lan}. This centrality range corresponds to $\etfcal>0.042$~\TeV\ in Pb+Pb collisions and $\etfcal>0.03$~\TeV\ in Xe+Xe collisions.
 
Charged-particle tracks~\cite{Aad:2015wga} are reconstructed from hits in the ID, and are subsequently used to construct the primary vertices.
Tracks are required to have $\pT>0.5$~\GeV\ and $|\eta|<2.5$ in Pb+Pb collisions and $\pT>0.3$~\GeV\ and $|\eta|<2.5$ in Xe+Xe collisions. They are required to satisfy the ``loose'' selection criteria, which require at least one pixel hit, with the additional requirement of a hit in the first pixel layer when one is expected, and at least six SCT hits. In addition, the distances of closest approach of the track to the primary vertex in both the transverse and longitudinal directions, $|d_0|$ and $|z_0\sin\theta|$, are required to be less than 1.5~mm~\cite{ATLAS:2011ah}. For evaluation of systematic uncertainties, the ``tight'' selection criteria are used, which further require at least two pixel hits, eight SCT hits, no missing hits in the pixel or SCT layers, and $|d_0|$ and $|z_0\sin\theta|$ values both smaller than 1~mm.  The primary particles in this analysis are defined as those with a lifetime larger than 30 picoseconds.  The influence of secondary particles originating from weak decays and detector material interactions are accounted for by varying track selection criteria and efficiency correction~\cite{ATL-INDET-INT-2010-017}.
 
The efficiency, $\epsilon(\pT,\eta)$, of the track reconstruction and track selection criteria is evaluated using Pb+Pb and Xe+Xe Monte Carlo events produced with the $\HIJING$ event generator~\cite{Gyulassy:1994ew}. The generated particles in each event are rotated in azimuthal angle according to the procedure described in Ref.~\cite{Masera:2009zz} to produce a harmonic flow consistent with the previous ATLAS measurements~\cite{ATLAS:2011ah,Aad:2012bu}. The response of the detector is simulated using $\GEANT$~\cite{Agostinelli:2002hh,Aad:2010ah} and the resulting events are reconstructed with the same algorithms as applied to the data. At mid-rapidity ($|\eta|<1$), the efficiency for central Pb+Pb collisions is about 67\% at 0.5~\GeV\ and increases to 71\% at higher $\pT$, while the efficiency for central Xe+Xe collisions is about 61\% at 0.3~\GeV\ and increases to 73\% at higher $\pT$~\cite{Aad:2019xmh}. For $|\eta|>1$, the efficiency decreases to about 40–60\% depending on the $\pT$ and centrality. The rate of falsely reconstructed (``fake'') tracks is also estimated and found to be significant only at $\pT<1$~\GeV\ in central collisions, where it ranges from 2\% for $|\eta|<1$ to 8\% at larger $|\eta|$. The fake rate drops rapidly for higher $\pT$ and for more peripheral collisions. The fake rate is accounted for in the tracking efficiency correction following the procedure in Ref.~\cite{Aad:2014vba}.
 
\section{Observables}
\label{sec:4}
In the experimental analysis, the measurement of $\rho_n$ in Eq.~\eqref{eq:rho} requires a calculation of the individual components in the numerator and denominator. For this purpose, Eq.~\eqref{eq:rho} is re-expressed as
\begin{align} \label{eq:rho1}
\rho_n =\frac{\cov{n}}{\sqrt{\var{n}}\sqrt{c_{k}}}\;,\;\cov{n} = \llrr{v_n^2\delta \pT}\;,\;\var{n}=\lr{v_n^4}-\lr{v_n^2}^2\;,\; c_k = \llrr{\delta\pT\delta\pT}\;.
\end{align}
The covariance $\cov{n}$ is a three-particle correlator, which is obtained by averaging over unique triplets in each event and then over all events in an event-activity class based on $\nchrec$ or $\etfcal$:
\begin{align}
\cov{n} &= \lr{\frac{\sum\nolimits_{i,j,k, i\neq j\neq k} w_iw_jw_k {\textrm e}^{{\textrm i}n(\phi_i-\phi_j)}(\ptk{k}-\lr{ [\pT]})}{\sum\nolimits_{i,j,k, i\neq j\neq k} w_iw_jw_k}}. \nonumber
\end{align}
In the above formula, the indices $i$, $j$, and $k$ loop over distinct charged particles to account for all unique triplets. The particle weight $w$ is constructed to correct for both detector nonuniformities and tracking inefficiency as explained in Section~\ref{sec:5}. The above expression for $\cov{n}$ can be expanded algebraically within the cumulant framework~\cite{Bilandzic:2010jr,Bilandzic:2013kga,Jia:2017hbm,Huo:2017nms} into a polynomial function of flow vectors and momentum-scalar quantities,
\begin{align}\label{eq:2}
&\bq_{n;k}=\frac{\sum_i w_i^k {\textrm e}^{{\textrm i}n\phi_i}}{\sum_i w_i^k}\;,\;p_{m;k}=\frac{\sum_i w_i^k(p_{\mathrm{T},i}-\lr{[\pT]})^m}{\sum_i w_i^k}\;,\; [\pT]=\frac{\sum_i w_i p_{\mathrm{T},i}}{\sum_i w_i}\;,
\end{align}
with $k$ and $m$ being integer powers. Details of this expansion can be found in Ref.~\cite{Zhang:2021phk}.
 
In order to reduce short-range nonflow correlations from resonance decays and jets, pseudorapidity gaps are often explicitly required between the particles in each triplet. This analysis uses the standard, two-subevent, and three-subevent methods to explore the influence of nonflow correlations~\cite{Jia:2017hbm,Zhang:2021phk}. In the standard method, all charged particles within $|\eta|<2.5$ are used. In the two-subevent method, triplets are constructed by combining particles from two subevents labeled $a$ and $c$, separated by a $\Deta$ gap in between to reduce nonflow effects: $-2.5<\eta_a<-0.75\;,\; 0.75<\eta_c<2.5$. The two particles contributing to the flow vector are chosen as one particle each from $a$ and $c$, while the third particle providing the $\pT$ is taken from either $a$ or $c$. In the three-subevent method, three nonoverlapping subevents $a$, $b$, and $c$ are chosen as $-2.5<\eta_a<-0.75\;,|\eta_b|<0.5\;,\; 0.75<\eta_c<2.5$. The particles contributing to flow are chosen from subevents $a$ and $c$ while the third particle is taken from subevent $b$.
 
In the large collision systems considered in this analysis, the nonflow effects are important only in peripheral collisions, where the statistical uncertainties of $\cov{n}$ are also large. The $\cov{n}$ values obtained from the two- and three-subevent methods agree within a few percent in mid-central and central collisions but show some differences in the peripheral region. In that region, however, only $\cov{2}$ has enough statistical precision to detect differences between the two- and three-subevent methods. Therefore, the final results for $\cov{2}$ are obtained from the three-subevent method, while the final results for $\cov{3}$ and $\cov{4}$ are obtained using triplets from both the two- and three-subevent methods, referred to as the ``combined-subevent'' method.
 
The statistical uncertainty of the measurement is obtained using a standard Poisson bootstrap method commonly employed in cumulant analyses~\cite{efron1979,PUB-2021-011}. Thirty pseudo-datasets were generated by assigning to each event a random Poisson weight with a mean of one, and the quantities in Eq.~\eqref{eq:rho1} are calculated for each pseudo-dataset. This method is mathematically justified when the number of events in a given event-activity class is sufficiently large. The standard deviations of the thirty values for each quantity were taken as the statistical uncertainties in the final result.
 
To obtain the Pearson coefficient in Eq.~\eqref{eq:rho1}, one also needs to calculate the variances $c_k$ and $\var{n}$. The former, $c_k=\lr{\sum\nolimits_{i,j, i\neq j} w_iw_j(\ptk{i}-\lr{[\pT]})(\ptk{j}-\lr{[\pT]})/\sum\nolimits_{i,j,i\neq j} w_iw_j}$, is obtained using all the pairs in the full event, i.e.\ within $|\eta|<2.5$. The latter is calculated in terms of the two-particle cumulant $c_n\{2\}\equiv\lr{v_n^2}$ and four-particle cumulant $c_n\{4\}\equiv\lr{v_n^4}-2\lr{v_n^2}^2$~\cite{Borghini:2000sa},
\begin{align} \label{eq:4}
\var{n} =  c_n\{4\}_{\mathrm{standard}} + c_n\{2\}^2_{\textrm{two-sub}}\;. \nonumber
\end{align}
The $c_n\{4\}$, being four-particle correlators, are known to be relatively insensitive to nonflow correlations but usually have poor statistical precision~\cite{Bilandzic:2010jr}. Therefore, they are obtained from the standard method in the full event. On the other hand, the two-particle cumulants $c_n\{2\}$ are more susceptible to nonflow correlations but at the same time provide better statistical precision. Therefore, the $c_n\{2\}$ are calculated from the two-subevent method with the $\eta$ choices discussed above. The calculation of $c_n\{2\}$ and $c_n\{4\}$ follow the procedure used in previous analyses~\cite{Bilandzic:2010jr,Jia:2017hbm}, i.e.\ they are expressed in terms of flow vectors $\bq_{n;k}$ defined in Eq.~\eqref{eq:2}.
 
The default $\eta$ ranges for the standard and subevent methods discussed above are listed in Table~\ref{tab:1}. In addition to these default values, the analysis is also repeated for $\eta$ ranges that are closer to mid-rapidity in order to study the impact of nonflow and longitudinal dynamics. This choice, listed in Table~\ref{tab:1} as ``Alternative $\eta$ selection'', could also be useful when comparing the results of this analysis with other experiments.
 
The charged particles used in this analysis are selected from some predefined $\pT$ ranges similar to those in a previous measurement~\cite{Aad:2019fgl}. For the analysis of Pb+Pb data, two ranges, $0.5<\pT<5$~\GeV\ and $0.5<\pT<2$~\GeV, are used. For the analysis of Xe+Xe data, one additional range with a lower threshold, $0.3<\pT<2$~\GeV\ is used. However, the primary results are based on the range $0.5<\pT<5$~\GeV, which has the best statistical precision.
 
\begin{table}[!h]
\centering
\caption{\label{tab:1} The $\eta$ and $\pT$ ranges chosen for the standard and subevent methods.}
\begin{tabular}{|l|c|c|}\hline
Method              &Default $\eta$ selection & Alternative $\eta$ selection\\\hline
Standard       & $|\eta|<2.5$                              &  $|\eta|<1$  \\
Two-subevent   &  $0.75<-\eta_a,\eta_c<2.5$    & $0.35<-\eta_a,\eta_c<1$ \\
Three-subevent  & $0.75<-\eta_a,\eta_c<2.5$,\;$|\eta_b|<0.5$  & $0.35<-\eta_a,\eta_c<1$,\;$|\eta_b|<0.3$ \\\hline
Combined-subevent  & \multicolumn{2}{|c|}{average of two-subevent and three-subevent results} \\\hline\hline\hline
\multicolumn{2}{|c|}{$\pT$ selection for Xe+Xe} & $\pT$ selection for Pb+Pb \\\hline
\multicolumn{2}{|c|}{$0.3<\pT<2$~\GeV, $0.5<\pT<5$~\GeV, $0.5<\pT<2$~\GeV}  &  $0.5<\pT<5$~\GeV, $0.5<\pT<2$~\GeV\\\hline
\end{tabular}
\end{table}
 
\section{Analysis procedure}
\label{sec:5}
 
The measurement of $\cov{n}$, $\var{n}$, and $c_k$ follows a procedure similar to that detailed in Refs.~\cite{Aaboud:2017blb,Aad:2019fgl} that consists of three steps. In the first step, these correlators are calculated in each event as the average over all combinations among particles from an $\eta$ range and a $\pT$ range listed in Table~\ref{tab:1}. In the second step, the values obtained in each event are averaged over events with comparable multiplicity, defined as events with either similar $\etfcal$ values ($|\Delta\etfcal|<$ 0.002~\TeV) or the same $\nchrec$. They are then combined in broader multiplicity ranges of the event ensemble to obtain statistically more precise results. The Pearson coefficients $\rhon{n}$ are then obtained via Eq.~\eqref{eq:rho1}. This event-averaging procedure is necessary to reduce the effects of centrality fluctuations for each event-activity estimator~\cite{Zhou:2018fxx,Schenke:2020uqq,Bozek:2020drh}.
 
In the third step, the $\nchrec$ dependence is converted to a centrality percentile dependence for each observable~\cite{Aad:2019fgl}. This is accomplished by calculating the average $\etfcal$ for each given $\nchrec$, which is then mapped to the centrality percentile. The mapping procedure is necessary such that results obtained for $\etfcal$ and $\nchrec$ can be directly compared using a common $x$-axis.
 
In order to account for detector inefficiencies and nonuniformities, particle weights defined in Section~\ref{sec:4} are calculated as
\begin{eqnarray}\label{eq:w}
w(\phi,\eta,\pT) = d(\phi,\eta)/\epsilon(\eta,\pT)\;.
\end{eqnarray}
The additional weight factor $d(\phi,\eta)$ accounts for nonuniformities in the azimuthal acceptance of the detector as a function of $\eta$ and amounts to a 5--20\% variation. To determine it, all reconstructed charged particles for a given $\pT$ selection are entered in a two-dimensional histogram $N(\phi,\eta)$, and the weight factor is then obtained as $d(\phi,\eta) \equiv \lr{N(\eta)}/N(\phi,\eta)$, where $\lr{N(\eta)}$ is the track density averaged over $\phi$ in the given $\eta$ bin. This procedure removes most $\phi$-dependent nonuniformity from the track reconstruction~\cite{Aaboud:2017acw}, and the resulting flow vectors ${\bq}_{n;k}$ in Eq.~\eqref{eq:2} should ideally be uniformly distributed in azimuthal angle. Any residual offsets are then subtracted by an event-averaged offset ${\bq}_{n;k} -\lr{{\bq}_{n;k}}_{\mathrm{evts}}$~\cite{Aad:2014fla}, which was implemented as an improvement over the previous measurement~\cite{Aad:2019fgl}.
 
\section{Systematic Uncertainties}
\label{sec:6}
The systematic uncertainties in this analysis arise from track selection, reconstruction efficiency, acceptance reweighting and centrality selection, and are evaluated for each observable: $\cov{n}$, $\var{n}$, $c_k$ and $\rhon{n}$. Systematic uncertainties from many sources enter the analysis through the particle weights in Eq.~\eqref{eq:2}. The uncertainties partially cancel out between the numerator and denominator in constructing $\rhon{n}$. The uncertainties quoted below are for the 0--50\% centrality range, and they are generally comparable between Xe+Xe and Pb+Pb. In the peripheral collisions, systematic uncertainties are smaller than the statistical uncertainties; they are evaluated but not quoted below because $\rhon{n}$ values are often very close to zero, and quoting the uncertainties as percentages is not very meaningful. The uncertainty contributions from different sources are described below with a focus of their impact on $\rhon{n}$.
 
From previous measurements~\cite{Aad:2012bu}, the $v_n$ signal has been shown to have a strong dependence on $\pT$ but relatively weak dependence on $\eta$. Therefore, a $\pT$-dependent uncertainty in the track reconstruction efficiency $\epsilon(\eta,\pT)$ could affect the measured signal through the particle weights. The uncertainties in the track reconstruction efficiency are due to differences in the detector conditions and known differences in the material between data and simulations. The uncertainties in detector efficiency vary in the range 1--4\%, depending on $\eta$ and $\pT$~\cite{ATLAS:2018ezv,Aad:2019xmh}. The systematic uncertainties for each observable are evaluated by repeating the analysis with the tracking efficiency increased and decreased by its corresponding uncertainty. The resulting uncertainties in $\rhon{n}$ are less than 2\% for $n=2$ and 4, but increase to 6\% for $n=3$ because the $\cov{3}$ values decrease towards zero in the peripheral region.
 
The contamination from fake tracks varies with the tracking selection. To assess how the fake tracks change the results, the requirements imposed on the reconstructed tracks are varied from those in the default track selection. The loose and tight track selections discussed in Section~\ref{sec:3} are used for this purpose. The differences are largest in the most central and peripheral collisions, where the correlation signals are smaller and the influence of fake tracks is thus higher. In Pb+Pb collisions, the differences are up to 3\%, 6\%, and 9\% for $n = 2$, 3, and 4, respectively. The uncertainties are smaller in Xe+Xe collisions due to lower rates of fake tracks, except in peripheral collisions for $n=3$ and 4.
 
The effect of detector azimuthal nonuniformity is accounted for by the weight factor $d(\eta,\phi)$ in Eq.~\eqref{eq:w}. The effect of reweighting is studied by setting the weight to one and repeating the analyses with the residual offset correction still applied to the flow vectors. The unweighted results generally agree with the weighted results within 1--3\%, except for peripheral Xe+Xe collisions, where the uncertainties are larger.
 
The centrality definitions used to classify the events into centrality percentiles in the 0--84\% range have a 1\% uncertainty, due to an inefficiency in selecting minimum-bias collisions. The impact of this uncertainty is evaluated by varying the centrality interval definitions by $\pm 1$\%, and recalculating all the observables. The impact for all observables is small in central and mid-central collisions, but becomes the dominant uncertainty in the more peripheral region. This type of uncertainty is only used when results are presented as a function of centrality percentiles. The uncertainties are 0--3\% in central and mid-central collisions and increase to 2--8\% in the more peripheral collisions depending on the harmonic number $n$ and collision system.
 
The systematic uncertainties from the different sources described above are added in quadrature to give the total systematic uncertainty for each observable. These uncertainties for $\rhon{n}$ are summarized in Table~\ref{tab:2}. The relative uncertainties are larger in central and peripheral collisions, where the values of $\rhon{n}$ are small relative to their statistical uncertainties.
 
\begin{table}[h!]
\centering
\caption{Sources of systematic uncertainty in percentage on the measured $\rhon{n}$ values.}
\label{tab:2}
\vspace*{0.2cm}
\begin{tabular}{|l|l| c c c| c cc|}\hline
Centrality & Sources   & \multicolumn{3}{c|}{Pb+Pb}         & \multicolumn{3}{c|}{Xe+Xe}  \\\cline{3-8}
&           & $\rho_2$ [\%]&  $\rho_3$ [\%]&  $\rho_4$ [\%]  & $\rho_2$ [\%] &  $\rho_3$ [\%] &  $\rho_4$ [\%]\\  \hline
& Efficiency        & 0.9      & 2.2      & 1.5    & 0.5       & 1.4       & 1.2 \\
& Track quality       & 0.5      & 0.6      & 9~~~      & 0.6       &  4.9      & 0.7\\
0--10\% & $\phi$ nonuniformity& 0.5   &$<0.5$~~~~    & 0.9    & 1~~~         & $<0.5$~~~~    & 2.3\\
& Centrality        & $<0.5$~~~~   &$<0.5$~~~~    & 1.5    & $<0.5$~~~~    & $<0.5$~~~~    & 0.5\\
& Total             & 2~~~        & 3~~~        & 10~~~~~     & 2~~~         & 6~~~         & 3~~~ \\
\hline
& Efficiency        & 1.2      & 3~~~       & 0.9     & 1.0       & 3.2       & 0.8 \\
& Track quality       & 1.6      & 3.6     & 1.5     &$<0.5$~~~~     & 2.3       & 2.5\\
20--30\% &  $\phi$ nonuniformity& $<0.5$~~~~   & $<0.5$~~~~ & 0.6 & $<0.5$~~~~    & 1.5       &  $<0.5$~~~~ \\
& Centrality        & $<0.5$~~~~   & 3~~~        & 1.5    & $<0.5$~~~~    & 1.5       & 0.6\\
& Total             & 2~~~        & 6~~~        & 3~~~      & 2~~~         & 4~~~         & 3~~~ \\
\hline
& Efficiency        & 1.3       & 6~~~        & 0.8   & $<0.5$~~~    & 15~~~~~        & 15~~~~~ \\
& Track quality       & 1.9       & 7.5      & 1.1   & 1.2       & 15~~~~~        & 0.6 \\
40--50\% &  $\phi$ nonuniformity& $<0.5$~~~~ & $<0.5$~~~~   & $<0.5$~~~ & $<0.5$~~~~    &1.5        & 7~~~ \\
& Centrality        & 1.4      & 1.5       & 0.8   & 1.8       & 6~~~         & $<0.5$~~~~ \\
& Total             & 3~~~        & 10~~~~~        & 2~~~     & 3~~~         & 22~~~~~        & 18~~~~~ \\
\hline
\end{tabular}
\centering
\end{table}
\section{Results}\label{sec:7}
The values of $c_k$, $\var{n}$, and $\cov{n}$ are calculated and combined to obtain $\rhon{n}$ for each choice of $\pT$ and $\eta$ ranges in Pb+Pb and Xe+Xe collisions as defined in Table~\ref{tab:1}.  In each case, they can be obtained with the event-averaging procedure in intervals of either $\etfcal$ or $\nchrec$ and those intervals are translated to average centrality values. The default event-averaging procedure is based on $\etfcal$. As described in Section~\ref{sec:4},  the primary results shown are calculated for charged particles in the range $0.5<\pT<5$~\GeV, using the three-subevent method for $\rhon{2}$ and the combined-subevent method for $\rhon{3}$ and $\rhon{4}$.
 
\subsection{Dependence on method and collision systems}\label{sec:7.1}
\begin{figure}[h!]
\centering
\includegraphics[width=0.97\linewidth]{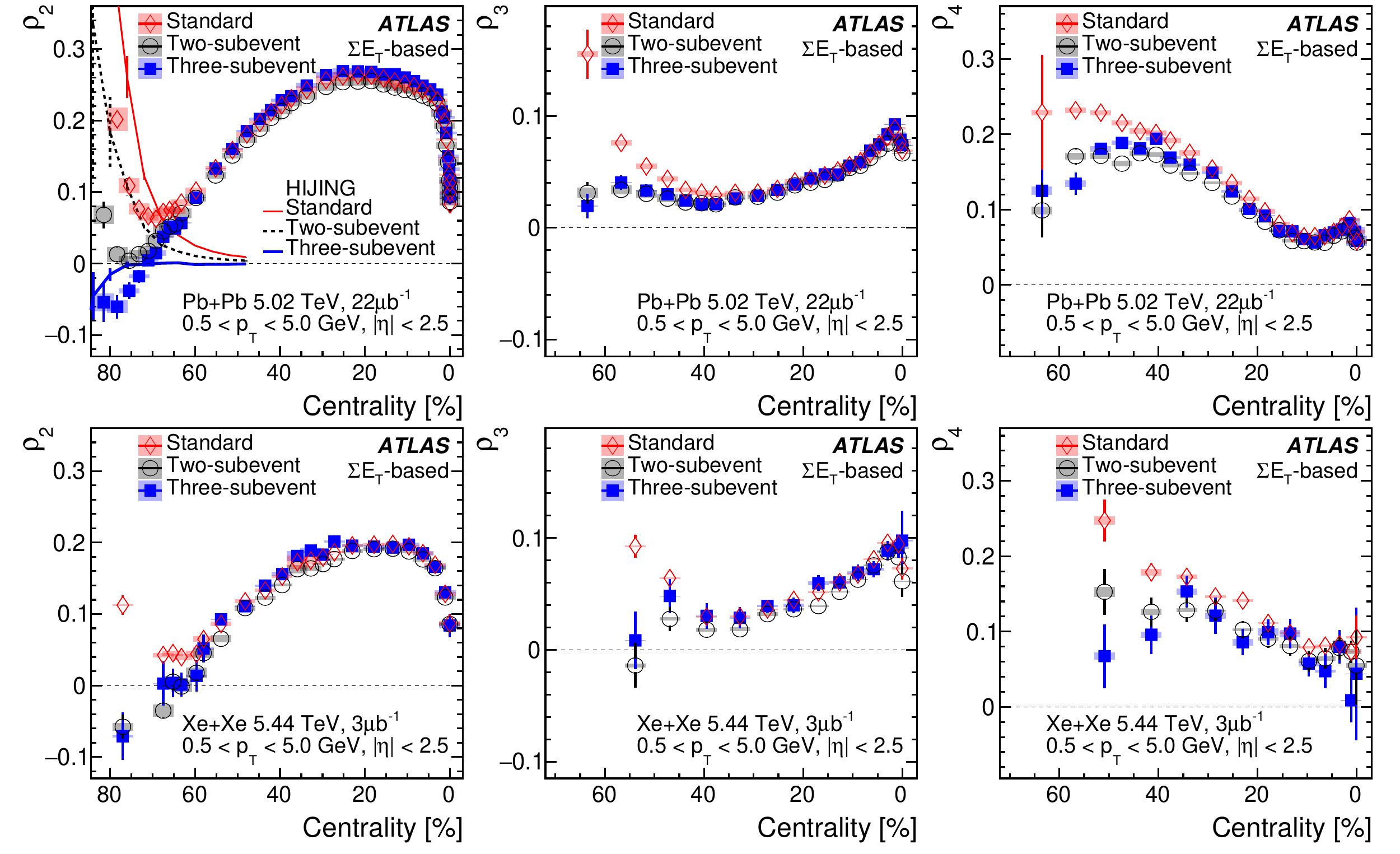}
\caption{The centrality dependence of $\rhon{n}$ for $n=$2 (left), 3 (middle), and 4 (right) in Pb+Pb (top) and Xe+Xe (bottom) collisions calculated for the standard, two-subevent, and three-subevent methods. They are calculated using the event-averaging procedure based on $\etfcal$. The error bars and shaded boxes represent statistical and systematic uncertainties, respectively. To reduce the statistical fluctuations in the Xe+Xe data, wider centrality binning is used in the bottom row. The Pb+Pb $\rhon{2}$ data are also compared with $\HIJING$ calculations from Ref.~\cite{Zhang:2021phk}, which includes only nonflow correlations.}
\label{fig:1}
\end{figure}
 
Figure~\ref{fig:1} shows the $\rhon{n}$ values obtained from the standard, two-subevent, and three-subevent methods for charged particles with $0.5<\pT<5$~\GeV\ in Pb+Pb and Xe+Xe collisions. They are obtained using the event-averaging procedure based on $\etfcal$ and plotted as a function of centrality. The results are close to each other in central and mid-central collisions. In peripheral collisions beyond 60\% centrality, the values from the standard method are significantly larger than those obtained from the subevent methods. This is consistent with the significant nonflow correlations arising from resonance decays and jets, which give positive contributions to both $v_n$ and $[\pT]$ in the standard method. The nonflow effects in the two-subevent method, reflected by the difference from the three-subevent method, are also visible beyond 70\% centrality. Smaller differences, albeit weakly dependent on centrality, are also observed between the two-subevent method and the three-subevent method in mid-central and central collisions. These differences are expected because the $v_n$ signal, as well as the decorrelations of $v_n$ and $[\pT]$, depend on the chosen $\eta$ intervals and $\Deta$, which differ between the two methods~\cite{Jia:2014ysa,Pang:2015zrq,Bozek:2016yoj}.
 
The influence of nonflow effects was investigated recently in models~\cite{Zhang:2021phk,Lim:2021auv}. The $\rhon{2}$ obtained from the $\HIJING$ model, which generates only nonflow correlations, shows a similar ordering between the three methods, as seen in Figure~\ref{fig:1}. In particular, the values of $\rhon{2}$ from the three-subevent method are closer to zero in the multiplicity range corresponding to the centrality range shown in Figure~\ref{fig:1}. The $\rhon{2}$ signal in the peripheral region cannot be reproduced by the $\HIJING$ model, which only includes nonflow correlations.

The results from the subevent methods show similar centrality dependences between the Pb+Pb and Xe+Xe: the $\rhon{2}$ values reach a minimum in the peripheral collisions, increase to a positive maximum value and then decrease in the most central collisions; the $\rhon{3}$ values show a mild increase towards central collisions; the $\rhon{4}$ values show an increase then a gradual decrease towards central collisions.
 
In the ultracentral collision region, all the $\rhon{n}$ show a sharp decrease towards the most central collisions. This decrease is much clearer in the Pb+Pb system due to its superior statistical precision and better centrality resolution than in the Xe+Xe system. This sharp decrease starts at around 1.6\% in centrality in Pb+Pb, which matches approximately to the location of the knee in the minimum-bias $\etfcal$ distribution~\cite{Aaboud:2019sma}. For events having $\etfcal$ values beyond the knee, essentially all nucleons participate in the collision. As a result, geometric fluctuations that enhance the $\rhon{n}$ values are suppressed. A similar suppression of fluctuations has also been observed for other flow observables~\cite{Aaboud:2019sma}.
 
\begin{figure}[h!]
\vspace*{-0.2cm}\centering
\includegraphics[width=0.97\linewidth]{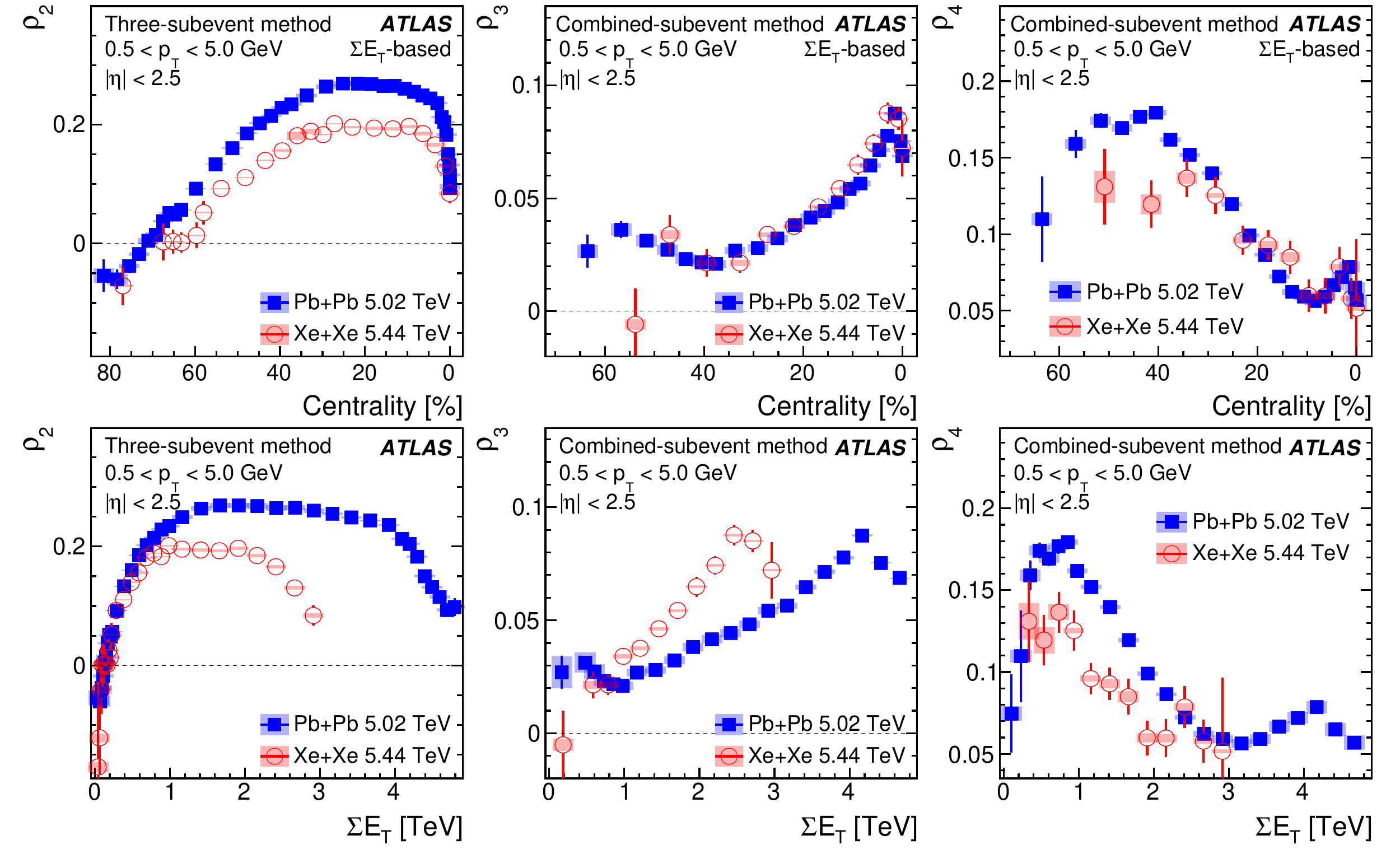}
\vspace*{-0.2cm}\caption{The centrality (top) and $\etfcal$ (bottom) dependences of $\rhon{n}$ for $n=2$ (left), 3 (middle) and 4 (right) in Pb+Pb and Xe+Xe collisions. They are calculated using the event-averaging procedure based on $\etfcal$. The error bars and shaded boxes represent statistical and systematic uncertainties, respectively.}
\label{fig:2}
\end{figure}

Figure~\ref{fig:2} provides a direct comparison of the Pb+Pb and Xe+Xe $\rhon{n}$ values as a function of centrality (top) and $\etfcal$ (bottom). These two different choices for the $x$-axis test whether the system-size dependence of $\rhon{n}$ scales with centrality or event multiplicity. When compared at the same centralities, the Xe+Xe $\rhon{2}$ values are everywhere smaller than the Pb+Pb values. However, when compared using  $\etfcal$, the Pb+Pb and Xe+Xe $\rhon{2}$ values agree for small $\etfcal$ values ($\etfcal<0.5$~\TeV) but differ for larger $\etfcal$. When plotted as a function of $\etfcal$, the $\rhon{3}$ values in Pb+Pb and Xe+Xe collisions are similar only at low $\etfcal$, while they are similar over the full range when plotted as a function of centrality. The $\rhon{4}$ values for the two systems are similar when plotted as a function of centrality in the 0--40\% centrality range, but not when plotted as a function of $\etfcal$.
 
\subsection{Dependence on the $\pT$ and $\eta$ ranges}\label{sec:7.2}
\begin{figure}[h!]
\centering
\includegraphics[width=0.97\linewidth]{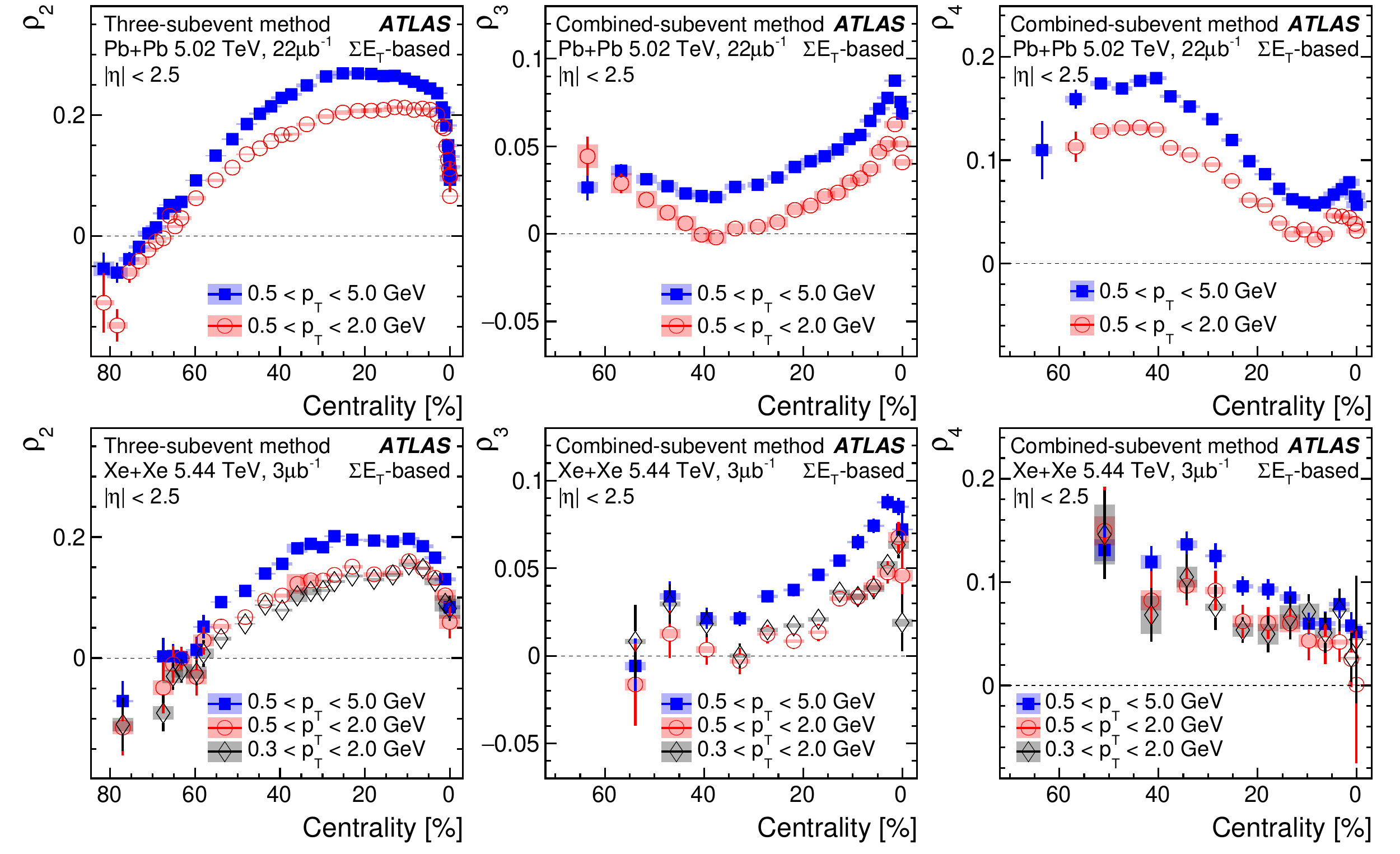}
\caption{The centrality dependence of $\rhon{n}$ for two $\pT$ ranges in Pb+Pb collisions (top) and three $\pT$ ranges in Xe+Xe collisions (bottom) for $n=$2 (left), 3 (middle) and 4 (right). They are obtained via the event-averaging procedure based on $\etfcal$. The error bars and shaded boxes represent statistical and systematic uncertainties, respectively.}
\label{fig:3}
\end{figure}

Figure~\ref{fig:3} shows the centrality dependence of $\rhon{n}$ in two $\pT$ ranges for Pb+Pb collisions and three $\pT$ ranges for Xe+Xe collisions. It is observed that the $\rhon{n}$ values for $0.5<\pT<2$~\GeV\ are smaller than those for $0.5<\pT<5$~\GeV\ in both systems, but the overall centrality dependence remains similar. In Xe+Xe collisions, the $\rhon{n}$ values obtained for a lower $\pT$ range of $0.3<\pT<2$~\GeV\ are found to be close to those obtained for $0.5<\pT<2$~\GeV. This is expected since the collective behavior of the bulk particles in the $0.5<\pT<2$~\GeV\ range reflects mainly hydrodynamic response, so including more particles by further lowering the $\pT$ threshold does not significantly change the $\rhon{n}$. This is an important observation for comparison with other experiments or model calculations, where different $\pT$ ranges are often used.

\begin{figure}[h!]
\centering
\includegraphics[width=0.97\linewidth]{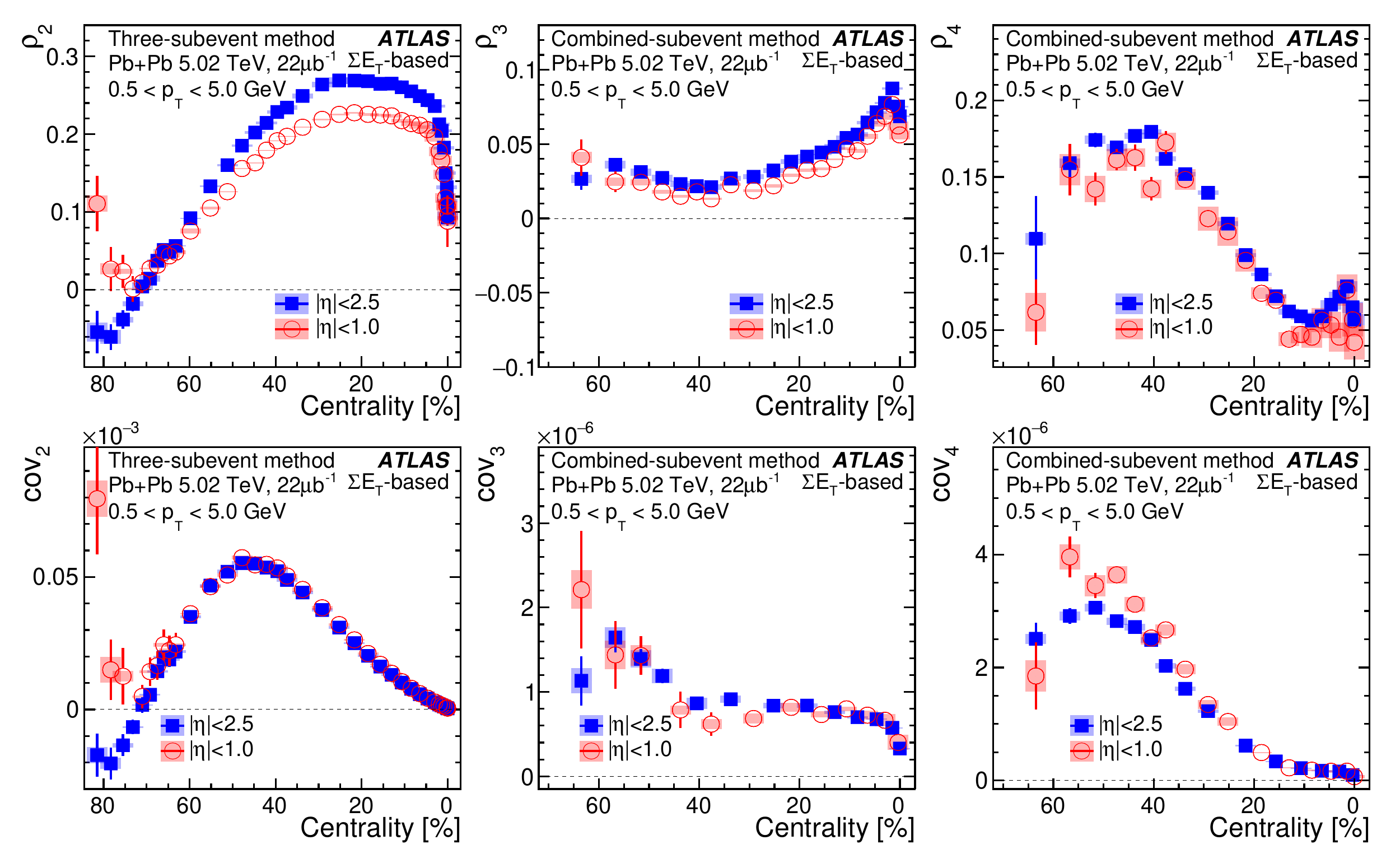}
\vspace*{-0.0cm}
\caption{The centrality dependence of $\rhon{n}$ (top) and $\cov{n}$ (bottom) for $n=$2 (left), 3 (middle), and 4 (right) in Pb+Pb collisions compared between the two choices for the $\eta$ ranges from Table~\ref{tab:1}. They are calculated using the event-averaging procedure based on $\etfcal$. The error bars and shaded boxes represent statistical and systematic uncertainties, respectively.}
\label{fig:4}
\end{figure}
 
The analysis is also repeated for the $\eta$ range closer to mid-rapidity, $|\eta|<1$, as listed in Table~\ref{tab:1}. Figure~\ref{fig:4} compares the centrality dependence of $\rhon{n}$ and $\cov{n}$ for the two $\eta$ ranges. The results for $\cov{n}$ are almost in agreement with each other, except for $n=2$ and 4 in peripheral collisions. In contrast, the results for $\rhon{n}$ are systematically lower for $|\eta|<1$ than for $|\eta|<2.5$. This implies that the difference arises from the $\eta$ dependence of the $\var{n}$ and $c_k$ values used to calculate $\rhon{n}$ via Eq.~\eqref{eq:2}. The values of $\cov{2}$ and $\rhon{2}$ for centrality above 70\% are larger for $|\eta|<1$, likely due to larger residual nonflow effects associated with a smaller $\eta$ range.
 
\subsection{Effects of centrality fluctuations}\label{sec:7.3}
\begin{figure}[h!]
\centering
\includegraphics[width=0.97\linewidth]{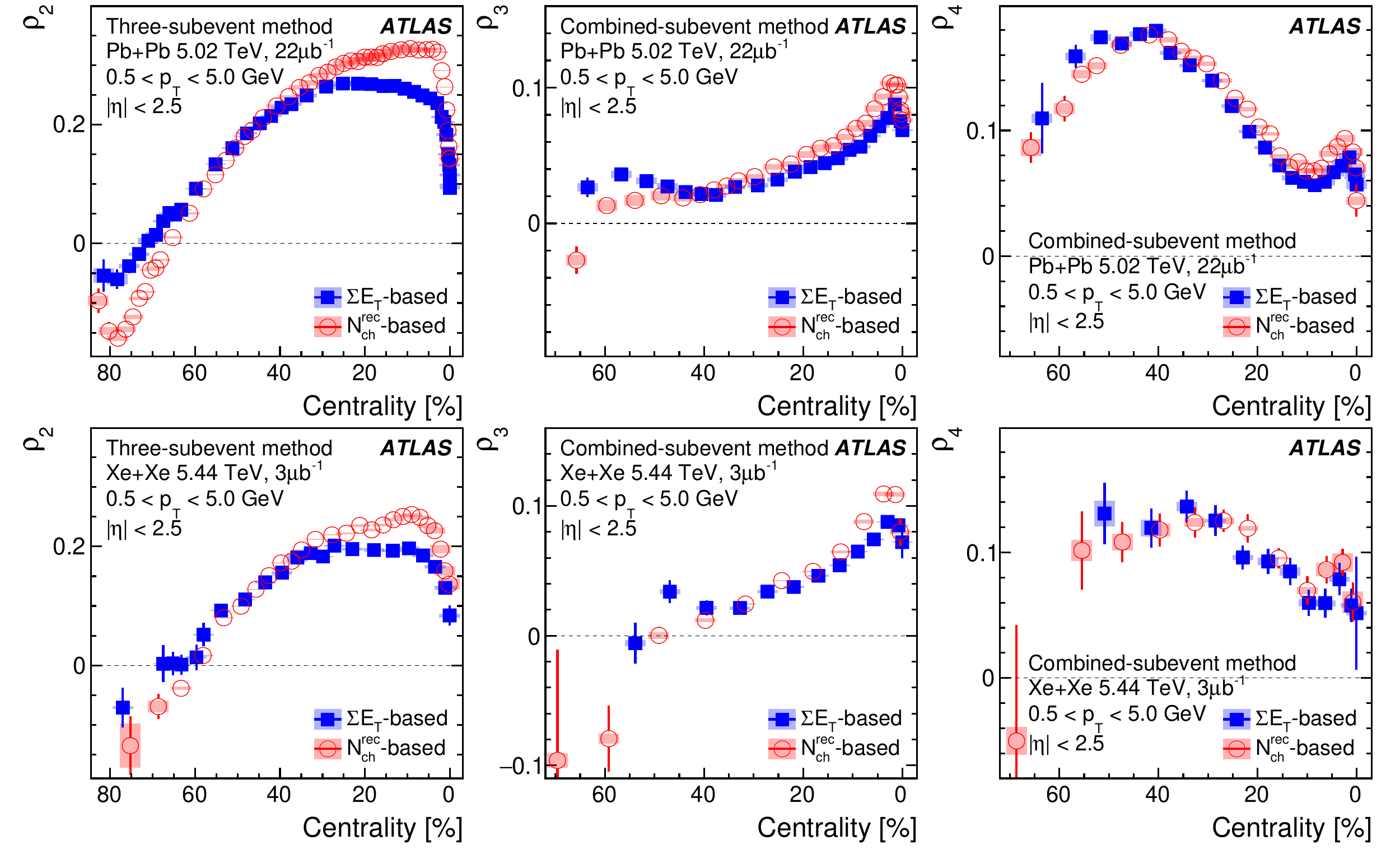}
\caption{The centrality dependence of $\rhon{n}$ in Pb+Pb (top) and Xe+Xe (bottom) collisions for $n=$2 (left), 3 (middle), and 4 (right), compared between the $\nchrec$-based event-averaging procedure (solid squares) and the $\etfcal$-based event-averaging procedure (solid circles). The results are calculated using charged particles with $0.5<\pT<5$~\GeV. The error bars and shaded boxes represent statistical and systematic uncertainties, respectively.}
\label{fig:5}
\end{figure}

As discussed in the introduction, due to the finite resolution of an event-activity estimator used to characterize the event centrality, the multiparticle cumulants for flow and $[\pT]$ fluctuations are sensitive to the multiplicity observable used in the event-averaging procedure. The results for $\rhon{n}$ as a function of centrality in Pb+Pb and Xe+Xe collisions are shown in Figure~\ref{fig:5}. Large differences between the $\rhon{2}$ values are observed in central collisions and in peripheral collisions: compared to results based on $\etfcal$, the results based on $\nchrec$ are larger in central collisions (0--40\% range) and smaller in peripheral collisions (beyond 50\% centrality). Differences between the two event activities are also observed for $\rhon{3}$ and $\rhon{4}$.
 
The influence of centrality fluctuations on $\rhon{n}$ was recently studied in a transport model framework~\cite{Jia:2021wbq}, albeit at RHIC energies of $\sqn=0.2$~\TeV. That study found that the $\rhon{2}$ values based on particle multiplicity at mid-rapidity are different from those based on particle multiplicity at forward rapidity. These differences are qualitatively similar to those observed in Figure~\ref{fig:5}. The $\rhon{2}$ values obtained using event activity at forward rapidities were also found to be more consistent with results obtained using the number of participating nucleons~\cite{Jia:2021wbq}. That finding reinforces the notion that the event-activity estimator in ATLAS based on $\etfcal$ has better centrality resolution than the estimator based on $\nchrec$.
 
Recently, it was argued that $\rhon{2}$ is a sensitive probe of the nature of collectivity in small collision systems and peripheral heavy-ion collisions, in particular for isolating the contribution from initial momentum anisotropy in a gluon saturation picture~\cite{Giacalone:2020dln}. The hydrodynamic expansion in the final state produces a negative (positive) $\rhon{2}$ in peripheral (nonperipheral) collisions~\cite{Bozek:2020drh,Giacalone:2020dln,Schenke:2020uqq}, while the initial momentum anisotropy is expected to give a large positive contribution in the most peripheral collisions~\cite{Giacalone:2020byk}. Therefore, the centrality dependence of $\rhon{2}$, after considering both the initial-state and final-state effects, is predicted to exhibit an increasing trend toward the most peripheral centrality~\cite{Giacalone:2020byk}. However, Figure~\ref{fig:5} shows that the trends of $\rhon{2}$ in peripheral collisions could still be modified by the centrality fluctuations.

\begin{figure}[h!]
\centering
\includegraphics[width=1\linewidth]{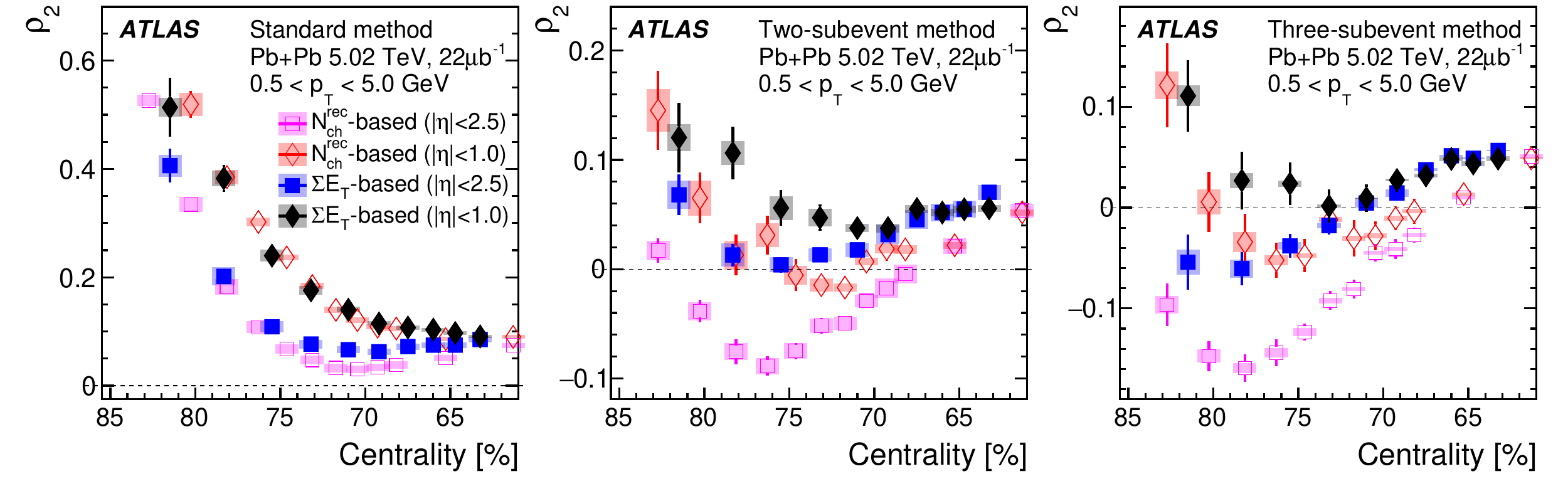}
\caption{ The centrality dependence of $\rhon{2}$ in Pb+Pb collisions in the peripheral region of 60--84\% for the standard method (left), two-subevent method (middle) and three-subevent method (right), compared between the $\nchrec$-based and  $\etfcal$-based event-averaging procedures and two $\eta$ ranges. The error bars and shaded boxes represent statistical and systematic uncertainties, respectively.}
\label{fig:6}
\end{figure}

Figure~\ref{fig:6} compares the centrality dependence of $\rhon{2}$ in $|\eta|<2.5$ and $|\eta|<1$ based on $\etfcal$ and $\nchrec$ in more detail over the 60--84\% centrality range. It is shown separately for the standard method and subevent methods in order to better separate the influence of nonflow effects from other effects. The successive reduction of the $\rhon{2}$ from the standard method in the left panel, to the two-subevent method in the middle panel, and to the three-subevent method in the right panel is a robust feature of suppression of the nonflow correlations~\cite{Zhang:2021phk}. In the right panel, where the residual nonflow is the smallest, two interesting features can be observed: 1) the $\rhon{2}$ values obtained for the narrow $|\eta|<1$ range are much larger than those for $|\eta|<2.5$, suggesting that the results obtained in $|\eta|<1$ still have significant nonflow contributions; 2) the differences between the $\rhon{2}$ values from the two event-activity estimators are large for both $\eta$ ranges, reflecting the impact of centrality fluctuations. The results from this measurement do not show clear evidence for initial-state momentum anisotropy. Future more detailed studies of the behavior of $\rhon{2}$ in very peripheral collisions, including smaller $pp$ and $p$+Pb collision systems, will be useful to disentangle the effects of nonflow, centrality fluctuations, and initial momentum anisotropy.

\section{Comparison with theory}\label{sec:8}
After the $\rhon{n}$ observable was proposed~\cite{Bozek:2016yoj}, several model predictions became available with different assumptions about the initial condition and final-state dynamics. Models that consider only the initial condition, such as Glauber or Trento models~\cite{Moreland:2014oya}, rely on a linear response relation between flow and eccentricity, $v_n\propto \varepsilon_n$, and between $[\pT]$ and the ratio of initial energy to initial entropy, $E/S$~\cite{Schenke:2020uqq,Giacalone:b}. From $\varepsilon_n$ and $E/S$, which can be calculated for each event without running the full hydrodynamic model simulation, the authors construct an estimator for the $\rhon{n}$ values. More realistic hydrodynamic models start from the Glauber or Trento model, and the system is evolved according to either two-dimensional (2D) boost-invariant or IP-Glasma full three-dimensional (3D) hydrodynamic equations. These models include the v-USPhydro model~\cite{Giacalone:2020dln}, the Trajectum model~\cite{Nijs:2021clz}, and the IP-Glasma+MUSIC model~\cite{Schenke:2020uqq,Giacalone:2020byk}. The first two are 2D hydrodynamic models based on a Trento initial condition and the third is a 3D hydrodynamic model based on a 3D initial condition that is dynamically generated from gluon saturation models. The latter has an option to include the contribution from initial momentum anisotropy ($\epsilon_p$). For comparison with the data, the predictions of IP-Glasma+MUSIC both with and without $\epsilon_p$ are included.  Most models have been tuned to describe reasonably well the bulk observables such as $v_2$, $v_3$ and $\pT$ spectra.  Many of these models also include the effects of nuclear quadrupole deformation with different values of the deformation parameters $(\beta,\gamma)$ in Eq.~\eqref{eq:R}. The chosen $\beta_{\mathrm{Xe}}$ value is 0.2 in Trento, 0.16 in Trajectum, and 0.18 in IP-Glasma+MUSIC. The chosen $\beta_{\mathrm{Pb}}$ value is 0.06 in Trento, but zero in other models. The default triaxiality value is chosen to be $\gamma_{\mathrm{Xe}}=30^{\circ}$ and $\gamma_{\mathrm{Pb}}=27^{\circ}$ in Trento and zero in other models. The choice of deformation parameter values in the Trento model is motivated by state-of-the-art nuclear energy-density functional calculations~\cite{Bally:2021qys,Budaca:2020ynk}, which predict the most probable values of $(\beta_{\mathrm{Xe}},\gamma_{\mathrm{Xe}})\approx(0.2,27^{\circ})$ and $(\beta_{\mathrm{Pb}},\gamma_{\mathrm{Pb}})\approx(0.06,27^{\circ})$. However, these calculations also predict that the shapes of Pb and Xe nuclei are not fixed, and can fluctuate over a broad range of $\gamma$ values.
 
\begin{figure}[h!]
\centering
\includegraphics[width=1\linewidth]{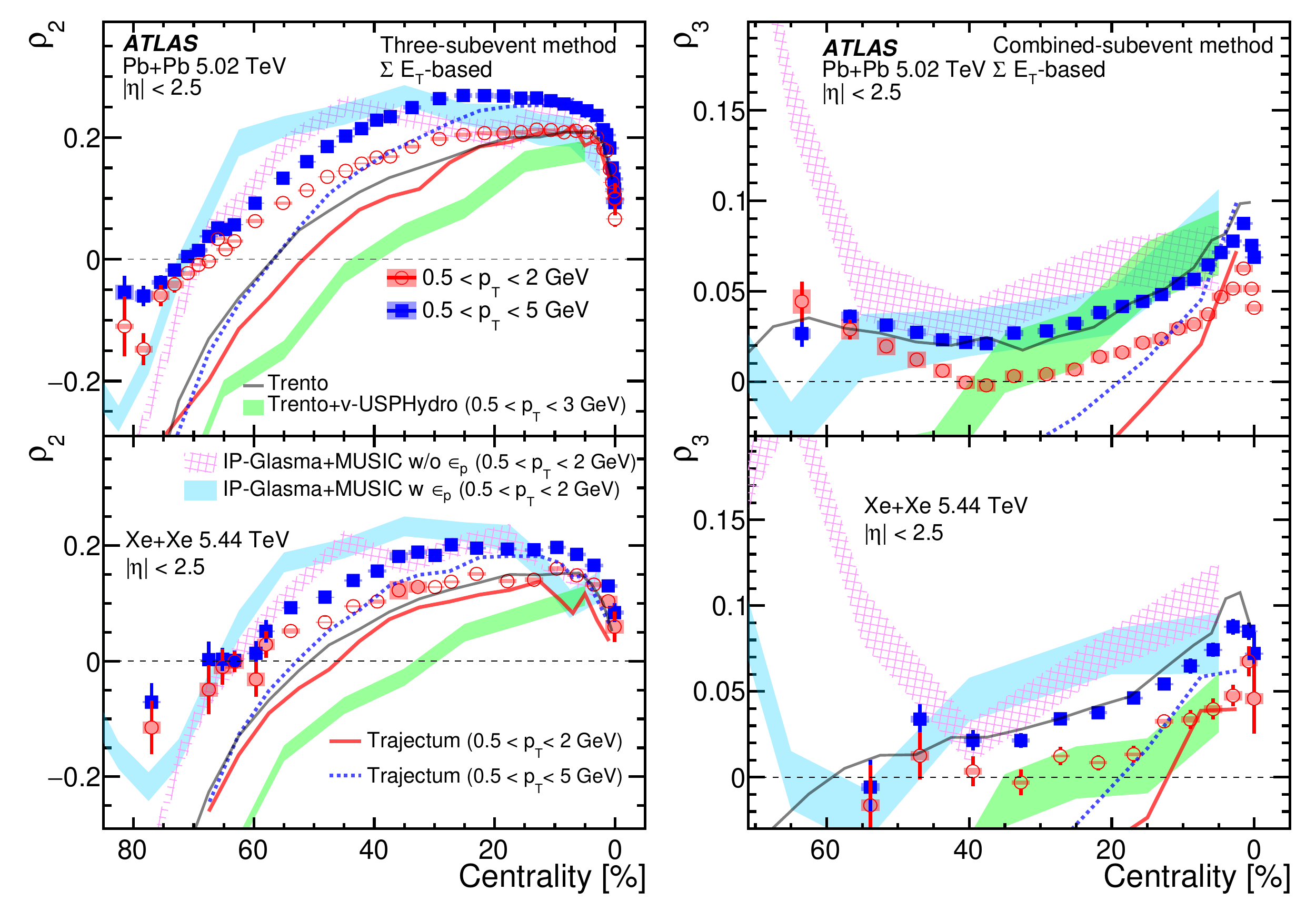}
\caption{The $\rhon{2}$ (left) and $\rhon{3}$ (right) values in Pb+Pb (top) and Xe+Xe (bottom) collisions in two $\pT$ ranges and $|\eta|<2.5$ compared with various models: Trento~\cite{Giacalone:b} and Trajectum~\cite{Nijs:2021clz} models in solid lines and v-USPhydro~\cite{Giacalone:2020dln} and IP-Glasma+MUSIC~\cite{Giacalone:2020byk} hydrodynamic models in shaded bands, which represent the statistical uncertainties of the model calculations. }
\label{fig:8}
\end{figure}
Figure~\ref{fig:8} shows the $\rhon{2}$ and $\rhon{3}$ values for two $\pT$ ranges in Pb+Pb (top) and Xe+Xe (bottom) collisions. They are compared with the various models described above.
 
In the 0--10\% centrality interval, where the effects of nuclear deformation are important, all models generally show reasonable agreement with each other and with the data. In particular, the Trajectum model quantitatively reproduces the ordering between $0.5<\pT<2$~\GeV\ and $0.5<\pT<5$~\GeV. In central Xe+Xe collisions, however, the Trajectum model underestimates the $\rhon{2}$ values, probably due to the smaller value of $\beta_{\mathrm{Xe}}$ used. In noncentral collisions, these models show significant differences from each other, which were recently shown to mainly reflect the different parameter values for the initial condition, in particular the nucleon size~\cite{Giacalone:2021clp}. In the peripheral collisions, all model predictions for $\rhon{2}$ show a sharp decrease and a sign-change, qualitatively consistent with the ATLAS data. The Trento model, which includes only initial-state effects, by construction has no $\pT$ or $\eta$ dependencies for $\rhon{2}$. The results from this model underestimate the values of $\rhon{2}$ in all $\pT$ ranges, describe the values of $\rhon{3}$ for $0.5<\pT<5$~\GeV, and overestimate the values of $\rhon{3}$ in other $\pT$ ranges. The v-USPhydro and Trajectum models significantly underestimate both the $\rhon{2}$ and $\rhon{3}$ values in noncentral collisions. IP-Glasma+MUSIC model predictions, both with and without $\epsilon_p$ included, are above the data in mid-central collisions (30--60\% centrality), but are below the data in more peripheral collisions. The IP-Glasma+MUSIC model predicts the location for the sign change but overestimates the data in mid-central collisions. The IP-Glasma+MUSIC model with $\epsilon_p$ shows differences from the model without $\epsilon_p$ in peripheral collisions beyond 70\% centrality, where the current data have limited precision. More detailed experimental measurements in that region, including smaller $pp$ and $p$+Pb collision systems, are needed to clarify the role of initial momentum anisotropy.

\begin{figure}[h!]
\centering
\includegraphics[width=0.97\linewidth]{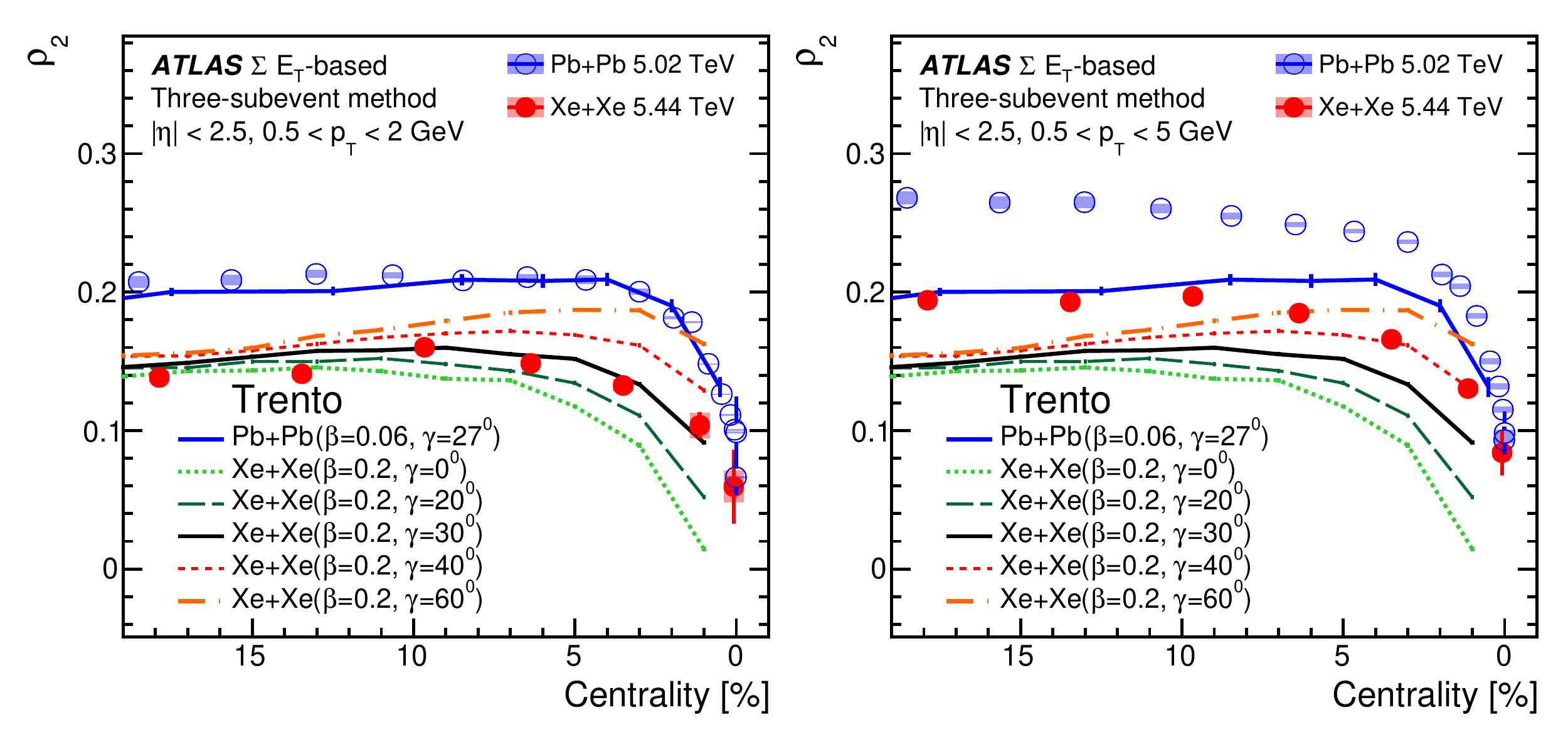}
\caption{Comparison of $\rhon{2}$ in Xe+Xe and Pb+Pb collisions with the Trento model for various quadrupole deformation parameter values~\cite{Bally:2021qys} in $0.5<\pT<2$~\GeV\ (left) and $0.5<\pT<5$~\GeV\ (right) as a function of centrality. The same Trento results are used in both panels, and they are connected by lines for better visualization.}
\label{fig:9}
\end{figure}

Figure~\ref{fig:9} compares $\rhon{2}$ data in the 0--20\% centrality range with the Trento model calculations to investigate the influence of triaxiality~\cite{Bally:2021qys}. Because of the large quadrupole deformation of the $^{129}$Xe nucleus, $\beta_{\mathrm{Xe}}\approx  0.2$, the $\rhon{2}$ should be sensitive to the triaxiality parameter $\gamma_{\mathrm{Xe}}$. This expectation is confirmed in the Trento model which produces significantly different trends for $\rhon{2}$ as a function of centrality for different $\gamma_{\mathrm{Xe}}$ values. However, comparisons between the Trento model and data require care as the $\pT$ dependence of $\rhon{n}$ is absent in the Trento model.
 
\begin{figure}[h!]
\centering
\includegraphics[width=0.8\linewidth]{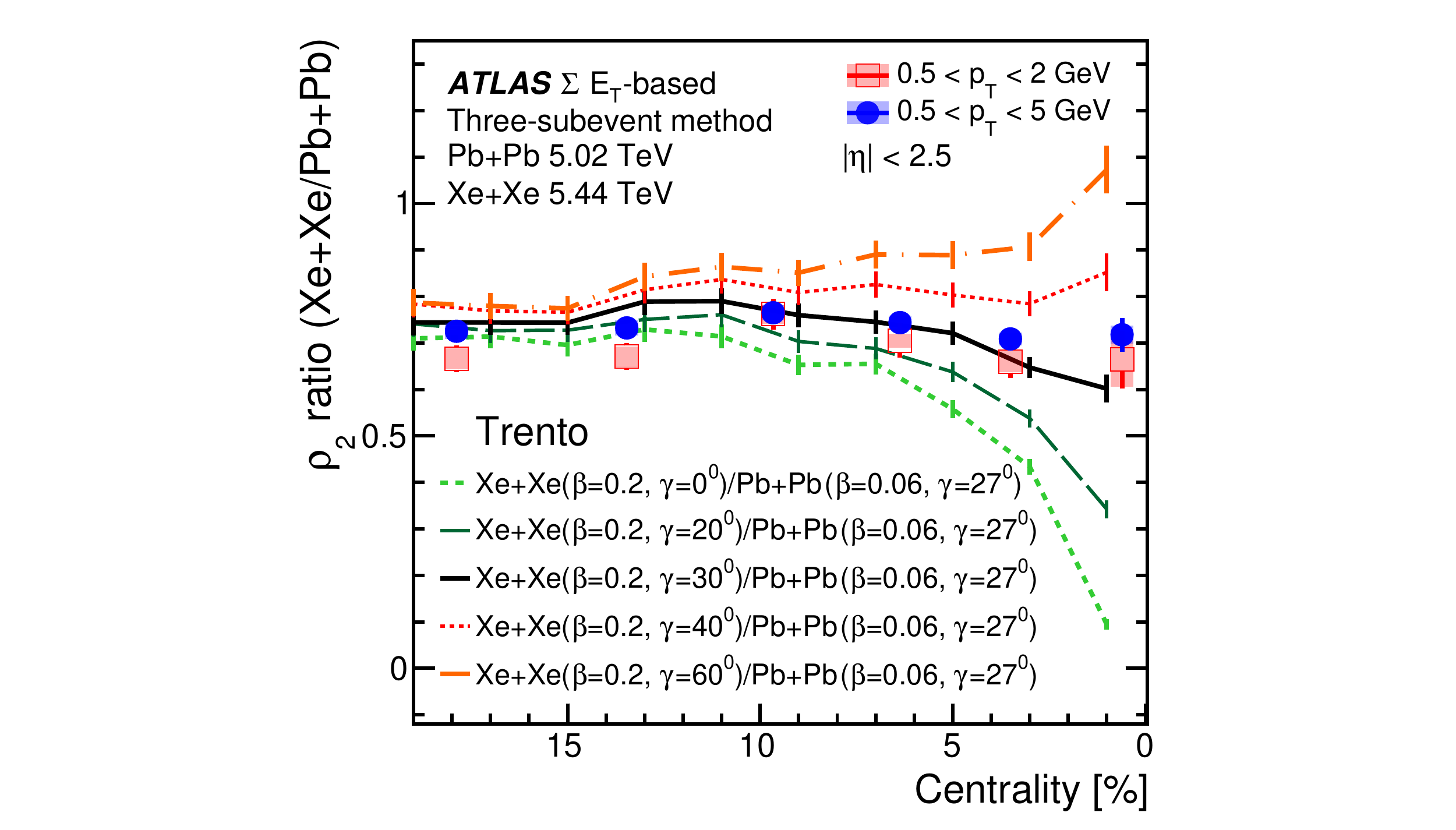}
\caption{Comparison of $\rho_{2}$ ratios, $\rho_{2,{\mathrm{Xe+Xe}}}/\rho_{2,{\mathrm{Pb+Pb}}}$, with the Trento model for various quadrupole deformation parameter values~\cite{Bally:2021qys} in two $\pT$ ranges.  The Trento model results are connected by lines for better visualization.}
\label{fig:10}
\end{figure}
 
In order to cancel out the $\pT$ dependence in the data, ratios of $\rhon{2}$ values between Xe+Xe and Pb+Pb are calculated for the two $\pT$ ranges and compared with the ratios obtained in the Trento model in Figure~\ref{fig:10}.  The ratio of the $\rhon{2}$ values is found to be approximately 0.7, and it is slightly lower in $0.5<\pT<2$~\GeV\ than in $0.5<\pT<5$~\GeV. In the 10--20\% centrality range, where the triaxiality plays a minor role, the model calculation is very close to the data ratio. In the 0--10\% centrality range, where the predicted $\rho_2$ values show significant dependence on the triaxiality, the comparison between the model and data favors a $\gamma_{\mathrm{Xe}}\approx 30^{\circ}$. This comparison provides clear evidence that flow measurements in central heavy-ion collisions can be used to constrain the quadrupole deformation, including the triaxiality, of the colliding nuclei. A future detailed comparison of the $\rhon{2}$ ratio with precision state-of-the-art hydrodynamic model calculations for different values of the deformation parameters would be needed to extract the most probable $\beta_{\mathrm{Xe}}$ and $\gamma_{\mathrm{Xe}}$ values.
 
\section{Conclusion}\label{sec:9}
 
This paper presents a measurement of the correlation between harmonic flow and the mean transverse momentum using 22 (470)~$\mu\textrm{b}^{-1}$ of minimum-bias (ultracentral) Pb+Pb collisions at $\sqn=5.02$~\TeV\ and 3~$\mu\textrm{b}^{-1}$ of minimum-bias Xe+Xe collisions at $\sqn=5.44$~\TeV\ recorded by the ATLAS detector at the LHC. The correlation between $v_n^2$ and the event-by-event $[\pT]$ is quantified using the Pearson correlation coefficient $\rhon{n}$, which is potentially sensitive to the shape and size of the initial geometry, nuclear deformation, and initial momentum anisotropy. Results are obtained for several $\pT$ and $\eta$ selections as a function of centrality, characterized by either $\nchrec$, the number of reconstructed charged particles in $|\eta|<2.5$, or $\etfcal$, the total transverse energy in the forward pseudorapidity range $3.2<|\eta|<4.9$. A comparison of results between these two event-activity estimators reveals the effects of centrality fluctuations.

The influence of nonflow contributions is studied using comparison between the standard, two-subevent, and three-subevent cumulant methods. The comparison between the three cumulant methods implies that the nonflow contribution has very little influence in the 0--70\% centrality range in both systems when using the subevent method. However, the results obtained from a smaller $\eta$ range, e.g.\ $|\eta|<1$, using the subevent method may still have significant nonflow contributions in the peripheral collisions. The results show significant differences between the two event-activity estimators. In central collisions, the $\rhon{n}$ values obtained with the $\nchrec$-based event-activity estimator are larger than those obtained with the $\etfcal$-based estimator for all harmonics, while the opposite trend is observed in peripheral collisions. These differences can be attributed to the relatively poorer centrality resolution associated with the $\nchrec$-based event-activity selection, which was also observed in previous flow measurements~\cite{Aaboud:2019sma}. Therefore, results based on $\etfcal$ are chosen as default results to be compared with models.
 
The results are compared with the Trento model calculation, which considers effects associated with the initial-state geometry and nuclear deformation, and three hydrodynamic models, v-USPhydro, Trajectum, and IP-Glasma+MUSIC, which consider the full space-time dynamics. The IP-Glasma+MUSIC model also has the option to include the contribution from the initial momentum anisotropy. All these models qualitatively describe the overall centrality- and system-dependent trends but fail to quantitatively reproduce all features of the data. In the peripheral collisions, the interpetation of $\rhon{2}$ in terms of initial momentum anisotropy is complicated by possible residual nonflow and centrality fluctuations. In the mid-central collisions, the IP-Glasma+MUSIC model overestimates the $\rhon{2}$ and $\rhon{3}$ values in data, while other models underestimate them. In the central collisions, most models show good agreement with the $\rhon{2}$ values in data. A comparison of the ratio $\rho_{2,\mathrm{XeXe}}/\rho_{2,\mathrm{PbPb}}$ with the Trento model implies that the $^{129}$Xe nucleus is a highly deformed triaxial ellipsoid, corresponding to a triaxiality value of $\gamma\approx 30^{\circ}$. This provides strong evidence of triaxial deformation of the $^{129}$Xe nucleus using high-energy heavy-ion collisions.
 
\section*{Acknowledgments}


We thank CERN for the very successful operation of the LHC, as well as the
support staff from our institutions without whom ATLAS could not be
operated efficiently.
 
We acknowledge the support of
ANPCyT, Argentina;
YerPhI, Armenia;
ARC, Australia;
BMWFW and FWF, Austria;
ANAS, Azerbaijan;
CNPq and FAPESP, Brazil;
NSERC, NRC and CFI, Canada;
CERN;
ANID, Chile;
CAS, MOST and NSFC, China;
Minciencias, Colombia;
MEYS CR, Czech Republic;
DNRF and DNSRC, Denmark;
IN2P3-CNRS and CEA-DRF/IRFU, France;
SRNSFG, Georgia;
BMBF, HGF and MPG, Germany;
GSRI, Greece;
RGC and Hong Kong SAR, China;
ISF and Benoziyo Center, Israel;
INFN, Italy;
MEXT and JSPS, Japan;
CNRST, Morocco;
NWO, Netherlands;
RCN, Norway;
MEiN, Poland;
FCT, Portugal;
MNE/IFA, Romania;
MESTD, Serbia;
MSSR, Slovakia;
ARRS and MIZ\v{S}, Slovenia;
DSI/NRF, South Africa;
MICINN, Spain;
SRC and Wallenberg Foundation, Sweden;
SERI, SNSF and Cantons of Bern and Geneva, Switzerland;
MOST, Taiwan;
TENMAK, T\"urkiye;
STFC, United Kingdom;
DOE and NSF, United States of America.
In addition, individual groups and members have received support from
BCKDF, CANARIE, Compute Canada and CRC, Canada;
PRIMUS 21/SCI/017 and UNCE SCI/013, Czech Republic;
COST, ERC, ERDF, Horizon 2020 and Marie Sk{\l}odowska-Curie Actions, European Union;
Investissements d'Avenir Labex, Investissements d'Avenir Idex and ANR, France;
DFG and AvH Foundation, Germany;
Herakleitos, Thales and Aristeia programmes co-financed by EU-ESF and the Greek NSRF, Greece;
BSF-NSF and MINERVA, Israel;
Norwegian Financial Mechanism 2014-2021, Norway;
NCN and NAWA, Poland;
La Caixa Banking Foundation, CERCA Programme Generalitat de Catalunya and PROMETEO and GenT Programmes Generalitat Valenciana, Spain;
G\"{o}ran Gustafssons Stiftelse, Sweden;
The Royal Society and Leverhulme Trust, United Kingdom.
 
The crucial computing support from all WLCG partners is acknowledged gratefully, in particular from CERN, the ATLAS Tier-1 facilities at TRIUMF (Canada), NDGF (Denmark, Norway, Sweden), CC-IN2P3 (France), KIT/GridKA (Germany), INFN-CNAF (Italy), NL-T1 (Netherlands), PIC (Spain), ASGC (Taiwan), RAL (UK) and BNL (USA), the Tier-2 facilities worldwide and large non-WLCG resource providers. Major contributors of computing resources are listed in Ref.~\cite{ATL-SOFT-PUB-2021-003}.


\printbibliography
\clearpage 
 
\begin{flushleft}
\hypersetup{urlcolor=black}
{\Large The ATLAS Collaboration}

\bigskip

\AtlasOrcid[0000-0002-6665-4934]{G.~Aad}$^\textrm{\scriptsize 101}$,
\AtlasOrcid[0000-0002-5888-2734]{B.~Abbott}$^\textrm{\scriptsize 119}$,
\AtlasOrcid[0000-0002-7248-3203]{D.C.~Abbott}$^\textrm{\scriptsize 102}$,
\AtlasOrcid[0000-0002-1002-1652]{K.~Abeling}$^\textrm{\scriptsize 55}$,
\AtlasOrcid[0000-0002-8496-9294]{S.H.~Abidi}$^\textrm{\scriptsize 29}$,
\AtlasOrcid[0000-0002-9987-2292]{A.~Aboulhorma}$^\textrm{\scriptsize 35e}$,
\AtlasOrcid[0000-0001-5329-6640]{H.~Abramowicz}$^\textrm{\scriptsize 150}$,
\AtlasOrcid[0000-0002-1599-2896]{H.~Abreu}$^\textrm{\scriptsize 149}$,
\AtlasOrcid[0000-0003-0403-3697]{Y.~Abulaiti}$^\textrm{\scriptsize 116}$,
\AtlasOrcid[0000-0003-0762-7204]{A.C.~Abusleme~Hoffman}$^\textrm{\scriptsize 136a}$
% ...
\bigskip
\\

% ...
$^{29}$Physics Department, Brookhaven National Laboratory, Upton NY; United States of America.\\
% ...
$^{35}$$^{(a)}$Facult\'e des Sciences Ain Chock, R\'eseau Universitaire de Physique des Hautes Energies - Universit\'e Hassan II, Casablanca;$^{(b)}$Facult\'{e} des Sciences, Universit\'{e} Ibn-Tofail, K\'{e}nitra;$^{(c)}$Facult\'e des Sciences Semlalia, Universit\'e Cadi Ayyad, LPHEA-Marrakech;$^{(d)}$LPMR, Facult\'e des Sciences, Universit\'e Mohamed Premier, Oujda;$^{(e)}$Facult\'e des sciences, Universit\'e Mohammed V, Rabat;$^{(f)}$Institute of Applied Physics, Mohammed VI Polytechnic University, Ben Guerir; Morocco.\\
% ...
$^{55}$II. Physikalisches Institut, Georg-August-Universit\"{a}t G\"ottingen, G\"ottingen; Germany.\\
% ...
$^{101}$CPPM, Aix-Marseille Universit\'e, CNRS/IN2P3, Marseille; France.\\
$^{102}$Department of Physics, University of Massachusetts, Amherst MA; United States of America.\\
% ...
$^{116}$Department of Physics, New York University, New York NY; United States of America.\\
% ...
$^{119}$Homer L. Dodge Department of Physics and Astronomy, University of Oklahoma, Norman OK; United States of America.\\
% ...
$^{136}$$^{(a)}$Departamento de F\'isica, Pontificia Universidad Cat\'olica de Chile, Santiago;$^{(b)}$Millennium Institute for Subatomic physics at high energy frontier (SAPHIR), Santiago;$^{(c)}$Instituto de Investigaci\'on Multidisciplinario en Ciencia y Tecnolog\'ia, y Departamento de F\'isica, Universidad de La Serena;$^{(d)}$Universidad Andres Bello, Department of Physics, Santiago;$^{(e)}$Instituto de Alta Investigaci\'on, Universidad de Tarapac\'a, Arica;$^{(f)}$Departamento de F\'isica, Universidad T\'ecnica Federico Santa Mar\'ia, Valpara\'iso; Chile.\\
% ...
$^{149}$Department of Physics, Technion, Israel Institute of Technology, Haifa; Israel.\\
$^{150}$Raymond and Beverly Sackler School of Physics and Astronomy, Tel Aviv University, Tel Aviv; Israel.\\
% ...
$^{a}$ Also Affiliated with an institute covered by a cooperation agreement with CERN.\\
$^{b}$ Also at Borough of Manhattan Community College, City University of New York, New York NY; United States of America.\\
$^{c}$ Also at Bruno Kessler Foundation, Trento; Italy.\\
$^{d}$ Also at Center for High Energy Physics, Peking University; China.\\
$^{e}$ Also at Centro Studi e Ricerche Enrico Fermi; Italy.\\
$^{f}$ Also at CERN, Geneva; Switzerland.\\
% ...

\end{flushleft}


\clearpage
 
\end{document}